\documentstyle[twocolumn,prb,aps,psfig,epsf,floats]{revtex} 
\begin{document}
\title{Exact duality and dual Monte-Carlo simulation
for the Bosonic Hubbard model}
\author{F. H\'ebert, G. G. Batrouni, H. Mabilat}
\address{
Institut Non-Lin\'eaire de Nice,
Universit\'e de Nice--Sophia Antipolis, 
1361 route des Lucioles,
06560 Valbonne, France}
%%%%%%%%%%%%%%%%%%%%%%%%%%%%%%%%%%%%%%%%%%%%%%%%%%%%%%%%%%%%%%%%%%
\address{\mbox{}}
\address{\parbox{14cm}{\rm \mbox{ } \mbox{ }
We derive the exact dual to the Bosonic Hubbard model. The dual
variables take the form of conserved current loops (local and
global). Previously this has been done only for the very soft core
model at very high density. No such approximations are made here. In
particular, the dual of the hard core model is shown to have a very
simple form which is then used to construct an efficient Monte Carlo
algorithm which is quite similar to the World Line algorithm but with
some important differences. For example, with this algorithm we can
measure easily the correlation function of the order parameter (Green
function), a quantity which is extremely difficult to measure with the
standard World Line algorithm. We demonstrate the algorithm for the
one and two dimensional hardcore Bosonic Hubbard models. We present
new results especially for the Green function and zero mode filling
fraction in the two dimensional hardcore model.  }}
\address{\mbox{}}
\address{\parbox{14cm}{\rm \mbox{ } \mbox{ }
}}
\maketitle

\section{Introduction}

Since its discovery by Kramers and Wannier~\cite{kramers}, duality has
continued to play an important role in physics. Originally, duality
related the ``high temperature'' phase of one system to the ``low
temperature'' phase of another. But it has long been realized that
duality is more general than what is encountered in thermal
statistical physics. In the case of quantum phase transitions, for
example, high and low temperatures are replaced by high and low
coupling constants which determine the size of quantum
fluctuations. As an example in the thermal case, the two dimensional
Ising model is self-dual {\it i.e.}  its low temperature phase is
mapped onto the high temperature phase of its dual which is also an
Ising model. This allowed Kramers and Wannier\cite{kramers} to
determine its exact critical temperature. The nature of the dual and
the question of self duality depend on several factors like the
symmetry group of the action, the type of model (spin, gauge etc) and
the dimensionality.\cite{kogut-savit} Duality has found many
applications.\cite{kogut-savit,batrouni} For example, a very
interesting recent application was to show that the bosonization of
fermionic systems in any number of dimensions can be effected using
the duality transformation.\cite{lutkin}

In quantum systems, such as the Bosonic Hubbard model, an {\it
approximate} duality transformation has been used to express the
partition function in terms of conserved electric current
loops.\cite{cha} This transformation is approximate in two ways: (1)
The boson density is assumed to be very high, in other words the model
is necessarily very soft core, and (2) the Villain form of the
action\cite{villain} is assumed. Because of this, the dual model thus
obtained is not even the dual of the Hubbard model but that of the
Quantum Phase Model (QPM). Still, this formulation was used very
productively to study various aspects of the ground state phase
diagram of the QPM numerically\cite{cha,otterlo} and
analytically.\cite{amico,igor} However, to study the bosonic Hubbard
model with no approximations (other than Trotter discretization), one
still had to use one of the other available algorithms. The World Line
algorithm\cite{worldline} proved very useful in one and two dimensions
with or without disorder.\cite{batrouni1,batrouni2} Improvements on
this (in the hard core case) came in the form of the loop (or cluster)
algorithm\cite{cluster} and the continuous time cluster
algorithm.\cite{cont-cluster} These two algorithms were designed to
eliminate long relaxation times (critical slowing down) at half
filling and in the absence of longer range interactions (like next
near neighbour). When these conditions are not met these algorithms
become extremely inefficient. In addition to questions of
inefficiency, a disadvantage shared by these algorithms is that even
though one can calculate, rather easily, the superfluid density and
other physical quantities such as the energy and
compressibility\cite{cha,otterlo,batrouni1,batrouni2,cluster,cont-cluster}
it is very difficult, if not impossible, to measure the correlation
function of the order parameter (the Green function, $G$) which is a
very interesting quantity.  The ``worm'' algorithm\cite{worms} can be
used to calculate $G$ but seems to converge very slowly in two
dimensions. The recently developed stochastic series expansion
method\cite{sandvik} is very promising. It overcomes the
inefficiencies of the loop algorithm, but whether it is efficient for
measuring $G$ is to be seen.

It is sometimes desirable to do the simulation in the canonical
ensemble especially when studying questions of phase separation. To
this end, we will first show here how to implement the duality
transformation {\it exactly} following the method of reference
~\cite{batrouni}. We will then construct a QMC algorithm valid in the
canonical and grand canonical ensembles and with or without
disorder. The efficiency of this algorithm does not deteriorate when
the system is doped away from half filling or in the presence of
longer range interactions. It also has the advantage of being able to
calculate, rather easily, $G$ in the grand canonical and canonical
ensembles.

The paper is organized as follows. In section II we show the duality
tranformation leaving the details for the appendix. In section III we
simplify the results and construct the QMC algorithm for hardcore
Bosons. In section IV we discuss some algorithmic improvements and
details and show tests of the algorithm in the one dimensional bosonic
Hubbard model in the hardcore limit. In section V we present new
simulation results for the two dimensional hardcore Bosonic Hubbard
model with near and next near neighbors. Section VI is for
conclusions.

\section{Duality transformation of the Bosonic Hubbard model}

\subsection{The Bosonic Hubbard model}

We are interested in simulating models whose Hamiltonian has the form

\begin{eqnarray}
H=-t\sum_{\langle
  r r'\rangle}(a_{r}^{\dagger}a_{r'}+a_{r'}^{\dagger}a_{r})
+ V_{0}\sum_{r} \frac{\hat n_{r} \left( \hat n_r -1 \right)}{2}
\\
+V_{1}\sum_{\langle rr' \rangle}\hat n_{r}\hat n_{r'}
+ V_{2}\sum_{\langle\langle rr'\rangle\rangle}\hat n_{r}\hat n_{r' } + \cdots
- \mu \sum_{r}\hat n_{r}.\nonumber
\end{eqnarray} 

\noindent In this equation, $t$ is the transfer integral (the hopping
parameter) which sets the energy scale (in the simulations we set it
equal to $1$), $\mu$ is the chemical potential, $V_0$, $V_1$, and
$V_2$ are respectively the contact, the near and the next near
neighbor interactions. The dots stands for other longer range
interactions that can be added.  On the square lattice, the next near
neighbor is chosen along the diagonal, while in one dimension the
choice is the obvious one.  $a_r$ and $a_r^{\dagger}$ are destruction
and creation operators on site $r$ satisfying the usual softcore boson
commutation laws, $[a_r,a_{r'}^{\dagger}]=\delta_{r,r'}$, and $\hat
n_r=a_r^{\dagger}a_r^{}$ is the number operator on site $r$.  $\langle
rr'\rangle$ and $\langle\langle rr'\rangle\rangle$ label sums over
near and next near neighbors. The first term which describes the
hopping of the bosons among near neighbor sites gives the kinetic
energy.

The quantum statistical mechanics of this system is given by the
partition function Z,

\begin{eqnarray}
Z & = & {\rm Tr } (e^{-\beta H}), \nonumber  \\
  & = & \sum_{\lbrace \psi \rbrace} \langle \psi | e^{-\beta H} | \psi \rangle, 
\end{eqnarray}

\noindent where $\beta$ is the inverse temperature and the sum runs over 
a complete set of states. The expectation value of a quantum operator
is given by

\begin{eqnarray} 
\langle {\cal O} \rangle = {1 \over Z} {\rm Tr}({\cal O}e^{-\beta
H}).
\end{eqnarray}

\subsection{Path integral representation}

We now need to find a c-number representation of the partition
function, which can then be incorporated into a simulation
algorithm. There are several such representations.

The occupation number representation is defined by the eigenstates $|
{\mathbf n} \rangle$ of the number operator $a^{\dagger}_r a_r$,

\begin{eqnarray}
| {\mathbf n}  \rangle = \left[ \prod_r \frac{
(a^{\dagger}_r)^{n_r}}{\sqrt{n_r!}}\right] | 0  \rangle, 
\end{eqnarray}

\noindent where   $| {\mathbf n}  \rangle$  stands for a complete set of
$n_r$, $| {\mathbf n} \rangle = |n_1,n_2,\cdots \rangle$ where $n_r$
is the number of bosons at site $r$. In this representation the
resolution of unity reads

\begin{eqnarray}
{\mathbf 1} = \sum_{{\mathbf n}} | {\mathbf n}  \rangle \langle {\mathbf n}|.
\end{eqnarray}

Another common representation uses coherent states $| \Phi \rangle $,
\textit{i.e.} eigenstates of the destruction operator,

\begin{eqnarray}
a_r| \Phi \rangle = \phi_r|\Phi \rangle, \nonumber\\
\langle  \Phi  | a_r^{\dagger} = \langle
 \Phi  | \phi^{\star}_r, 
\end{eqnarray}

\noindent where the eigenvalues $\phi_r$ are complex numbers
defined on the sites of the lattice.  We completely define this state
with the vector $\Phi = (\phi_1,\dots,\phi_r,\dots) $.

In terms of the occupation number representation vacuum $|0\rangle$
the coherent state $|{\Phi}\rangle$ is defined as

\begin{eqnarray}
|\Phi \rangle = \exp \left( \sum_r
\phi_ra_r^{\dagger} \right) |0\rangle,
\end{eqnarray}

\noindent so that its projection on an occupation number state is

\begin{eqnarray}
\langle {\mathbf n} | \Phi \rangle =
\prod_r \left(\frac{\phi_{r}^{n_{r}}}{\sqrt{n_{r}!}}   \right).
\end{eqnarray}

\noindent With this normalization we obtain the inner product of two
coherent states

\begin{eqnarray}
\langle  \Psi  | \Phi  \rangle  =  
\exp \left( \sum_r \psi^{\star}_r \phi_r \right),
\end{eqnarray}

\noindent and the resolution of unity

\begin{eqnarray}
{\mathbf 1} = \int \prod_r {{d^2\phi_r} \over {\pi}} 
 \exp\left( - \sum_r |\phi_r|^2 \right) | \Phi \rangle \langle 
\Phi |.
\end{eqnarray}

The interaction and chemical potential part of the hamiltonian are
more easily expressed in the occupation number representation while a
coherent state representation is well suited to the treatment of the
kinetic term. We will use both successively.  We write the partition
function, Eq.~2, as

\begin{eqnarray}
Z &=& \sum_{\{{\mathbf n}\}} \langle {\mathbf n}| e^{-\beta H}
|{\mathbf n} \rangle \nonumber\\
&=& \sum_{\{{\mathbf n}\}} \langle {\mathbf n}|e^{-\delta H} 
e^{-\delta H} \cdots e^{-\delta H} |{\mathbf n} \rangle,
\end{eqnarray}

\noindent where $\delta \equiv \beta/L_{\tau}$, and $L_{\tau}$ is a large
enough integer such that $\delta \ll 1$. We express the partition
function this way because we can now express the exponentials in
Eq.~11 in a form suitable for easy evaluation. Between each pair of
exponentials introduce the resolution of unity, Eq.~5. We find

\begin{eqnarray}
Z= \smash{\sum_{\{{\mathbf n}_{\tau}\}}} 
\langle {\mathbf n}_1| e^{-\delta H}
|{\mathbf n}_{L_{\tau}} \rangle
\langle {\mathbf n}_{L_{\tau}}| e^{-\delta H}
|{\mathbf n}_{L_{\tau}-1} \rangle 
\cdots& \nonumber  \\
\cdots \langle {\mathbf n}_3| e^{-\delta H}
|{\mathbf n}_2 \rangle
\langle {\mathbf n}_2| e^{-\delta H}
|{\mathbf n}_1 \rangle .&
\end{eqnarray}

As $\delta$ is small,

\begin{equation}
e^{-\delta H} = e^{-\delta (T+V)} \approx e^{-\delta V/2} 
e^{-\delta T} e^{-\delta V/2} + {\mathcal O}(\delta^3),
\end{equation}

\noindent where $T$ is the kinetic energy and $V$ contains the interaction 
and chemical potential terms.

The partition function now becomes

\begin{eqnarray}
Z &=& \smash{\sum_{\{{\mathbf n}_{\tau}\}}} 
\langle {\mathbf n}_1|e^{-\delta V/2}
e^{-\delta T} e^{-\delta V/2}|{\mathbf n}_{L_{\tau}} \rangle
\cdots \nonumber \\
&&\qquad \qquad \cdots
\langle {\mathbf n}_2|e^{-\delta V/2}
e^{-\delta T} e^{-\delta V/2}
|{\mathbf n}_1 \rangle \nonumber \\
&=&  \smash{\sum_{\{{\mathbf n}_{\tau}\}}\prod_\tau} \left[ e^{-\delta V(
{\mathbf n}_{\scriptstyle \tau})} \right]\nonumber \\ 
&&\qquad \qquad \langle {\mathbf n}_1|
e^{-\delta T} |{\mathbf n}_{L_{\tau}} \rangle
\cdots \langle {\mathbf n}_2|
e^{-\delta T} 
|{\mathbf n}_1 \rangle  .
\end{eqnarray}

\noindent The error introduced by neglecting commutators of $T$ 
and $V$ is of order $L_\tau \delta^3 = \beta \tau^2$ for the partition
function.

We now introduce unity operators expressed in terms of coherent states
(Eq.~10) around the kinetic terms.

\begin{eqnarray}
Z&=&  \sum_{\{{\mathbf n}_{\tau}\}}
\int \prod_{r,\tau} \left(\frac{d^2\phi_{r,\tau} 
d^2\tilde{\phi}_{r,\tau}}{\pi^2}
e^{\textstyle{-(|\phi_{r,\tau}|^2 + |\tilde{\phi}_{r,\tau}|^2)}}\right) \nonumber \\
&\times&\prod_\tau \left[ e^{-\delta V(
{\mathbf n}_{\scriptstyle \tau})} \right]\langle {\mathbf n}_1 | \tilde{\Phi}_1 \rangle 
\langle \tilde{\Phi}_1 | e^{-\delta T} | \Phi_{L_{\tau}} \rangle
\langle \Phi_{L_{\tau}} | {\mathbf n}_{L_{\tau}} \rangle \nonumber \\ 
&\times&\langle {\mathbf n}_{L_{\tau}} | \tilde{\Phi}_{L_{\tau}} \rangle
\langle \tilde{\Phi}_{L_{\tau}} | e^{-\delta T} | \Phi_{L_{\tau}-1} \rangle
\langle \Phi_{L_{\tau}-1} | {\mathbf n}_{L_{\tau}-1} \rangle \nonumber \cdots \\
&\times&\langle {\mathbf n}_2 | \tilde{\Phi}_2 \rangle \langle \tilde{\Phi}_2 | e^{-\delta T}
| \Phi_1 \rangle \langle \Phi_1 | {\mathbf n}_1 \rangle .  
\end{eqnarray}

\noindent Notice that we need to introduce two sets of $L_\tau$ coherent 
states, the $\Phi$ and $\tilde{\Phi}$.  We can avoid introducing two
different representations (occupation and coherent) and only use
coherent states. But then, we would have to decouple the interaction
terms in order to integrate them out. At the end this would lead to
the same results, but it would be more difficult to recover a simple
expression for $Z$.
  
At this stage, it is customary to neglect terms of order 2 and higher
in $\delta$ in the expectation value of $\exp(-\delta T)$ since their
contribution goes to zero with $\delta$.  While this is correct and
would still allow us to calculate thermodynamic quantities such as
$\langle E \rangle$, it would not be sufficient to calculate the Green
function, $G(|r-r'|)=\langle a_r a_{r'}^{\dagger}\rangle$, at
distances $|r - r'| > 1$ (see below).

For this reason, it is very important to keep higher orders. We start
by keeping all such higher order terms of $\exp(-\delta T)$.  Since
$T$ is quadratic

\begin{eqnarray}
T = \sum_{r r'} a^{\dagger}_r K^{r r'} a_{r' } 
= {\mathbf a^{\dagger} K a},
\end{eqnarray}

\noindent where ${\mathbf  K }$ is a c-number matrix, we can use the 
following identity\cite{negele}

\begin{eqnarray}
& &\langle \Psi | e^{-\delta {\mathbf a^{\dagger} K a}} | \Phi \rangle
\nonumber \\ &=& \langle \Psi | :\exp\left(- {\mathbf a^{\dagger}}
\left[ 1 - e^{-\delta {\mathbf K}} \right]{\mathbf a} \right): | \Phi
\rangle,
\end{eqnarray}

\noindent where $ :{\mathcal O}: $ indicates that the operator 
${\mathcal O}$ is normal ordered. Eq.~17 may now be expressed as
\cite{negele}

\begin{eqnarray}
& &\langle \Psi | e^{-\delta {\mathbf a^{\dagger} K a}} | \Phi \rangle
\nonumber \\ && \qquad = \langle \Psi | \Phi \rangle \exp\left(-
\Psi^{\dagger} \left[ 1 - e^{-\delta {\mathbf K}} \right] \Phi
\right), \nonumber \\ && \qquad = e^{\Psi^{\dagger}\Phi} \exp\left(-
\Psi^{\dagger} \left[ 1 - e^{-\delta {\mathbf K}} \right] \Phi
\right).
\end{eqnarray}

Expanding the exponential term in Eq.~18, one finds

\begin{eqnarray}
e^{\Psi^{\dagger}\Phi} \exp - \biggl( \Psi^{\dagger} \biggl[ \delta
{\mathbf K} &-& \frac{\delta^2 {\mathbf K}^2}{2!}  \nonumber \\ &+&
\frac{\delta^3 {\mathbf K}^3}{3!}  - \frac{\delta^4 {\mathbf K}^4}{4!}
+ \cdots \biggr] \Phi \biggr) .
\end{eqnarray}

For simplicity, we will illustrate how to proceed by using one
dimensional $L_x$-site lattice with periodic boundary conditions. All
nearest neighbors $\langle r r' \rangle$ pairs are labeled by choosing
$r' = r + 1$ and summing over all $r$.  The same can be easily done in
any dimensionality. We write ${\mathbf K}$ as ${\mathbf K} = -t
{\mathbf K}_1$ where ${\mathbf K}_1$ is the matrix that connects
nearest neighbors: If $r$ and $r'$ are nearest neighbors then 
$K_{1}^{rr'} = 1$, otherwise $K_{1}^{r r'} = 0$. It is easy to show that

\begin{eqnarray}
{\mathbf K}^2 &=& (-t)^2 \left( 2  {\mathbf I + K}_2
\right),\nonumber \\
{\mathbf K}^3 &=& (-t)^3 \left( 3 {\mathbf K}_1 + {\mathbf K}_3
\right), \nonumber \\
{\mathbf K}^4 &=& (-t)^4 \left( 6  {\mathbf I} + 4 {\mathbf K}_2 
+ {\mathbf K}_4 \right) \cdots 
\end{eqnarray}

\noindent where ${\mathbf K}_L$ generates hops of $L$ lattice
spacings.

Introducing this into Eq.~19 and keeping only the leading order term
in $\delta$ for each matrix, ${\mathbf K}_L$, we get

\begin{eqnarray}
\exp  \biggl( \Psi^{\dagger}  \biggl[ {\mathbf I } +  ( \delta t ){\mathbf K}_1
 &+& \frac{(\delta t)^2 {\mathbf K}_2}{2!}  \\ &+& \frac{(\delta
 t)^3{\mathbf K}_3}{3!}  + \frac{ (\delta t)^4 {\mathbf K}_4}{4!}  +
 \cdots \biggr] \Phi \biggr). \nonumber
\end{eqnarray}

Substituting the preceding expression into Eq.~15, using Eq.~8 to
express the scalar products, and neglecting higher order terms, we get
the following expression for the partition function

\begin{eqnarray}
Z  =  \sum_{\{{\mathbf n}_\tau\}}\int \prod_{r,\tau}
\left( \frac{d^2\phi_{r,\tau}d^2\tilde{\phi}_{r,\tau}}{\pi^2} \right) 
{\mathcal P}(\Phi,\tilde{\Phi},{\mathbf n}),
\end{eqnarray}

\noindent where the weight ${\mathcal P}$ is given by

\begin{eqnarray}
&&{\mathcal P}(\Phi,\tilde{\Phi},{\mathbf n}) =
\prod_\tau \left[ e^{-\delta V(
{\mathbf n}_{\tau})} \right]\\ &\times&\prod_{r,\tau} \Biggl[
e^{-(|\phi_{r,\tau}|^2 + |\tilde{\phi}_{r,\tau}|^2)}
\frac{\left(\tilde{\phi}_{r,\tau}
\phi^{\star}_{r,\tau}\right)^{n_{r,\tau}}}{n_{r,\tau}!}   
\times e^{\tilde{\phi}^{\star}_{r,\tau+1}\phi_{r,\tau}}
\Biggr]\nonumber\\
 &\times& \exp \left(\sum_{r,r',\tau} \left(
 \tilde{\phi}^{\star}_{r',\tau+1} \left[
\delta t K_1^{r'r} + \frac{(\delta t)^2 K_2^{r'r}}{2!} + \cdots \right]
\phi_{r,\tau} \right) \right). \nonumber
\end{eqnarray}

We reiterate that as $\delta t \rightarrow 0$, the contributions of
multiple hops, \textit{ i.e.} ${\mathbf K}_L$ where $L > 1$, to the
energy and other local physical quantities vanish since they are of
higher order in $\delta t$.  However, this is not true for non local
quantities such as $G(r)=\langle a_0 a^{\dagger}_r \rangle$ for $r >
1$. Keeping the higher order terms will give us the generating
functional for this correlation function. Notice, however, that we
have kept only the terms necessary to calculate each quantity to
leading order in $\delta$. Therefore the algorithm will have
systematic Trotter errors of order $\beta \delta$.

\subsection{Dual formulation}

In order to perform the integration over the original variables and
obtain the dual formulation of Z, we express the $\phi_{r,\tau}$ and
$\tilde{\phi}_{r,\tau}$ as,

\begin{equation} 
\phi_{r,\tau} = \rho_{r,\tau} \exp (i\theta_{r,\tau}), 
\end{equation} 

\noindent which gives the following expression for Z

\begin{eqnarray}
Z &=&\int {\mathcal D}\Psi \prod_{r,\tau} \Biggl[
\,\frac{1}{n_{r,\tau}!} \left(\tilde{\rho}_{r,\tau}
\rho_{r,\tau}\right)^{\textstyle{n_{r,\tau}}}
 e^{\textstyle{i(\tilde{\theta}_{r,\tau}-\theta_{r,\tau})n_{r,\tau} }}
 \nonumber \\ &\times& \exp \left( \tilde{\rho}_{r,\tau+1}
 \rho_{r,\tau} e^{\textstyle -i(\tilde{\theta}_{r,\tau+1} -
\theta_{r,\tau}) } \right) \nonumber \\
&\times& \exp \left( \delta t
\tilde{\rho}_{r+1,\tau+1} \rho_{r,\tau} e^{\textstyle -i
(\tilde{\theta}_{r+1,\tau+1} -
\theta_{r,\tau}) } \right) \\                     
&\times& \exp \left( \delta t
\tilde{\rho}_{r,\tau+1} \rho_{r+1,\tau} e^{\textstyle -i
(\tilde{\theta}_{r,\tau+1} -
\theta_{r+1,\tau}) } \right) \nonumber \\
&\times&  \exp \left( \frac{(\delta t_{r,\tau})^2}{2!}
\tilde{\rho}_{r+2,\tau+1} \rho_{r,\tau} e^{\textstyle -i
(\tilde{\theta}_{r+2,\tau+1} -
\theta_{r,\tau}) } \right) \nonumber \\
&\times&  \exp \left( \frac{(\delta t_{r,\tau})^2}{2!}
\tilde{\rho}_{r,\tau+1} \rho_{r+2,\tau} e^{\textstyle -i
(\tilde{\theta}_{r,\tau+1} -
\theta_{r+2,\tau}) } \right)\cdots \Biggl]. \nonumber 
\end{eqnarray}

\noindent Dots stand for all longer range hopping terms and 
$\int {\mathcal D}\Psi$ stands for

\begin{eqnarray}
\int {\mathcal D}\Psi &=& 
\sum_{\{n_{r,\tau}\}} \int \prod_{r,\tau} \Biggl[ \rho_{r,\tau} d\rho_{r,\tau}
\tilde{\rho}_{r,\tau} d\tilde{\rho}_{r,\tau} \frac{d\theta_{r,\tau}}{\pi}
\frac{d\tilde{\theta}_{r,\tau}}{\pi} \nonumber \\
&\times& e^{\textstyle{-(|\rho_{r,\tau}|^2 +|\tilde{\rho}_{r,\tau}|^2)
}} \Biggr] e^{\textstyle{- \sum_\tau \delta V({\mathbf n}_\tau)}}.
\end{eqnarray}

As expected, only gauge invariant phase differences appear in
Eq.~25. Therefore, instead of performing the integrals over the gauge
dependent site phases, $\theta_{r,\tau}$ and
$\tilde{\theta}_{r,\tau}$, we express the partition function in term
of the following gauge invariant quantities

\begin{eqnarray}
&&\Theta^0_{r,\tau} =\theta_{r,\tau} -\tilde{\theta}_{r,\tau}, \nonumber \\
&&\Theta^1_{r,\tau} =\theta_{r+1,\tau} -\theta_{r,\tau},\nonumber \\
&&\tilde{\Theta}^0_{r,\tau} =\tilde{\theta}_{r,\tau+1} -\theta_{r,\tau}, \nonumber \\
&&\tilde{\Theta}^1_{r,\tau} =\tilde{\theta}_{r+1,\tau} -\tilde{\theta}_{r,\tau}.
\end{eqnarray}

\noindent We use indeces 0 and 1 to label the imaginary-time and
$\hat{x}$ directions.  In higher dimension, we would have 2
corresponding to the $\hat{y}$ direction, 3 to $\hat{z}$...

This variable change results in a non-trivial Jacobian which was
derived in reference~3. It was shown \cite{batrouni} that this
Jacobian is only the product of two kinds of constraints on the gauge
variables and that it can be written as

\begin{eqnarray}
{\mathcal J} &&= \sum_{ \{ N^{\scriptstyle \mu}_{r,\tau}\}} {\sum_{ \{
J^{\scriptstyle \mu}_{r,\tau}\}}}' \prod_{r,\tau} \Biggl[
e^{\textstyle{i \Theta^0_{r,\tau} (J^0_{r,\tau-1} + N^0_{r}) }}
\nonumber \\ &&\qquad \qquad \times \ e^{\textstyle{i
\tilde\Theta^0_{r,\tau} (J^0_{r,\tau} + N^0_{r}) }}
\, e^{\textstyle{i \Theta^1_{r,\tau} (J^1_{r,\tau} + N^1_{\tau}) }} \Biggr],
\end{eqnarray}  

\noindent where the integer valued variables, $N^{\mu}_{r,\tau}$ 
and $J^{\mu}_{r,\tau}$, represent conserved currents and are, in fact,
the dual variables\cite{batrouni}.  The origin of these constraints is
that the sum of the bond variables along any directed closed path must
be zero as is seen from their definition in Eq.~27. $\tilde{\Theta}^1$
does not appear in the hamiltonian and so has been eliminated in the
Jacobian. For a detailed discussion of these constraints and their
relation to duality, the strong coupling expansion etc see Ref~3.

The global currents, $N^{\mu}_{r,\tau}$, traverse the system from one
end to the other and are required by the periodic boundary conditions.
$N^{0}_{r}$ is a global current in the time direction and, therefore,
does not depend on the coordinate $\tau$. Similarly, $N^{1}_{\tau}$,
the global current in the $\hat x$ direction depends only on $\tau$
(in two dimensions it would also depend on $y$). The local currents,
$J^{\mu}_{r,\tau}$, form topologically trivial closed loops,
\textit{i.e.}  they do not wrap around the system from one end to the
other. The sum of the $J^{\mu}_{r,\tau}$ configurations is restricted
to be only over conserved current configurations: The total current
entering a site equals the current leaving.  This is indicated by the
prime on the sum.  We will see below that this current has a simple
physical interpretation.

With this variable change and Jacobian, Z becomes

\begin{eqnarray}
Z &=&\int {\mathcal D}\Psi 
 \prod_{r,\tau} \Biggl[ \,\frac{1}{n_{r,\tau}!} \left(\tilde{\rho}_{r,\tau} 
\rho_{r,\tau}\right)^{\textstyle{n_{r,\tau}}}
 e^{\textstyle{-i \Theta^0_{r,\tau} n_{r,\tau}}} \nonumber \\ &\times&
 \exp \left( \tilde{\rho}_{r,\tau+1} \rho_{r,\tau} e^{\textstyle -i
\tilde{\Theta}^0_{r,\tau}) } \right)\\
&\times& \exp \left( \delta t
\tilde{\rho}_{r+1,\tau+1} \rho_{r,\tau} e^{\textstyle -i(\tilde{\Theta}^0_{r+1,\tau} +
\Theta^1_{r,\tau}) } \right) \nonumber \\                     
&\times& \exp \left( \delta t
\tilde{\rho}_{r,\tau+1} \rho_{r+1,\tau} e^{\textstyle -i(\tilde{\Theta}^0_{r,\tau} -
\Theta^1_{r,\tau}) } \right) \nonumber \\
&\times&  \exp \smash{\biggl( \frac{(\delta t_{r,\tau})^2}{2!}}
\vphantom{ \vrule height 15pt depth 0pt width 0pt}
\tilde{\rho}_{r+2,\tau+1} \rho_{r,\tau} 
\nonumber \\
&&\qquad \qquad \qquad e^{\textstyle -i(\tilde{\Theta}^0_{r+2,\tau} +
\Theta^1_{r+1,\tau} + \Theta^1_{r,\tau} ) } \smash{\biggr)} \nonumber \\
&\times& \exp \smash{\biggl( \frac{(\delta t_{r,\tau})^2}{2!}}
\vphantom{ \vrule height 15pt depth 0pt width 0pt}
\tilde{\rho}_{r,\tau+1} \rho_{r+2,\tau} \nonumber \\
&& \qquad \qquad \qquad e^{\textstyle -i(\tilde{\Theta}^0_{r,\tau} -
\Theta^1_{r,\tau} -\Theta^1_{r+1,\tau} ) } \smash{\biggr)}
\vphantom{ \vrule height 0pt depth 5pt width 0pt}\cdots  \nonumber \\
& \times &e^{\textstyle{i \Theta^0_{r,\tau} (J^0_{r,\tau-1} + N^0_{r})
}} e^{\textstyle{i \tilde\Theta^0_{r,\tau} (J^0_{r,\tau} + N^0_{r}) }}
\nonumber \\ && \qquad \qquad \qquad \qquad \qquad \quad
e^{\textstyle{i \Theta^1_{r,\tau} (J^1_{r,\tau} + N^1_{\tau}) }}
\smash{\Biggr]}, \nonumber
\end{eqnarray}

\noindent where     
\begin{eqnarray} 
\int {\mathcal D} \Psi &=& \sum_{ \{ N^{\scriptstyle \mu}_{r,\tau}\}} 
{\sum_{ \{ J^{\scriptstyle \mu}_{r,\tau}\}}}'
\sum_{\{n_{r,\tau}\}} \\
\int \prod_{r,\tau}&& \Biggl[ \rho_{r,\tau} \frac{d\rho_{r,\tau}}{\pi}
\tilde{\rho}_{r,\tau} \frac{d\tilde{\rho}_{r,\tau}}{\pi} d\Theta^0_{r,\tau}
d\tilde{\Theta^0}_{r,\tau}
d\Theta^1_{r,\tau} \nonumber \\
&\times& e^{\textstyle{-(|\rho_{r,\tau}|^2 +|\tilde{\rho}_{r,\tau}|^2) }} \Biggr]
e^{\textstyle{- \sum_\tau \delta V({\mathbf n}_\tau)}}. \nonumber
\end{eqnarray}

Integrating the original variables, $\Theta$, $\rho$, and $n$ leaves
only the integer valued dual variables. The details are shown in the
appendix.  The result of this integration is a complicated expression
for the dual partition function involving several integer valued
fields and constraints.

One can greatly simplify this expression by explicitly solving the
constraints appearing in it (these constraints come from the
integration of $\Theta$), that is, by finding the allowed
configurations of $N^{\mu}_{r,\tau}$ and $J^{\mu}_{r,\tau}$ and their
associated Boltzmann weight.  In the appendix we illustrate this for
softcore and hardcore bosons but concentrate on the hard core case
where the dual partition function takes a particularly simple form.

\section{Quantum Monte Carlo algorithm in the hard core limit}

\subsection{Partition function in the hardcore limit}

As shown in the appendix, we can express the partition function and
other physical quantities in terms of only one field, the total
conserved hard-core current $ N^{\mu}_{r,\tau} + J^{\mu}_{r,\tau}$. By
hard-core, we mean that the total current, $ N^{\mu}_{r,\tau} +
J^{\mu}_{r,\tau} $, traversing a bond in the imaginary time direction
takes the values $0$ or $1$, while in a spatial direction it takes the
values $0, \, \pm 1$.  We have also chosen, without any loss of
generality, to keep $\sum_{r,\tau}J^{\mu}_{r,\tau} = 0$. Notice that
the total time currents are never negative: The bosons cannot go
backward in the imaginary time direction.

The partition function in the grand canonical case reads

\begin{eqnarray}
 Z&& =
 \sum_{ \{J, N \}} {\mathcal P}(N,J), \\
 &&=\sum_{ \{J, N \}}\Biggl( \prod_L \biggl[ \left(\frac{(\delta t)^L}{L\,!}\right)
 ^{N_{\scriptstyle L}}\biggr] 
\nonumber \\
&& \qquad \qquad \times  \  e^{ \delta \mu \sum_{r,\tau}N^{\scriptstyle 0}_{\scriptstyle r}
 } 
 e^{ -\delta V_{\scriptstyle 1} N_{\scriptstyle V_1}}  
 e^{ -\delta V_{\scriptstyle2} N_{\scriptstyle V_2}} \Biggr),  \nonumber
\end{eqnarray}

\noindent where $N_{L}$ is the number of jumps of length $L$ on the lattice, $N_{V_1}$  
the number of near neighbor time currents, and  $N_{V_2}$ 
the number of next near neighbor time currents.

The chemical potential appears in an exponential of $\beta 
\sum_{r,\tau}N^0_{r,\tau} $. So the number of
particles in the system is simply  

\begin{eqnarray}
\langle \hat{n} \rangle  = \frac{1}{\beta} 
\frac{\partial}{\partial \mu} \ln Z = \left\langle \frac{1}{L_{\tau}} \sum_{r,\tau}
N^0_{r} \right\rangle = \left\langle \sum_{r} N^0_{r} \right\rangle,
\end{eqnarray} 

\noindent which demonstrates that the current simply represents the 
current of bosons crossing the lattice in the positive imaginary time
direction.  So we have current lines crossing the lattice and the
sampling of the configurations is achieved by deforming these lines.
If one wants to work in the canonical ensemble, one simply has to fix
the number of current lines.

The effect of interactions is simply given by multiplying the weight
by $e^{ - \delta V_{\scriptstyle 1}}$ ($e^{ - \delta V_{\scriptstyle
2}}$ ) for every pair of near neighbor (next near neighbor) time
currents.

The term containing $t$, coming from the expansion of the hopping
term, also has a simple physical interpretation: Kinetic energy arises
when there are jumps, \textit{i.e.}  spatial currents. This term can
also be written as

\begin{eqnarray}
\prod_L \biggl[ \left(\frac{1}{L!}\right)
 ^{N_{\scriptstyle L}}\biggr] \left(\delta t\right)^{N_H}.
\end{eqnarray}

\noindent The exponent $N_H$ of $\delta t$ is the number of non zero 
spatial currents $N_H = \sum_L N_L = \sum_{r,\tau} |N^1_{\tau} +
J^1_{r,\tau}|$ but there is a combinatorial prefactor arising from
multiple jumps.
\begin{figure}[t]
\epsfxsize=3.0in
\epsfysize=2.25in
\epsffile{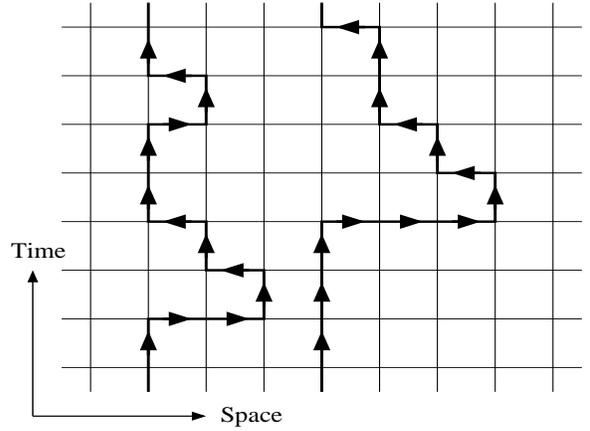}
\vskip 0.2cm
\caption{Example of a configuration of the conserved Boson current on 
a 1d lattice.}
\end{figure}

We therefore have a simple physically intuitive expression for $Z$
which allows easy evaluation of the Boltzmann weight.  For example in
Fig.~1, there are 7 single jumps, one double jump, and one triple
jump. There are also one pair of near neighbor time slices and one
pair of next near neighbor. So the weight associated with this
configuration is

\begin{eqnarray}
\frac{\left( \delta t \right)^{12}}{2! \, 3!} e^{-\delta V_1} e^{-\delta V_2}.
\end{eqnarray}

\subsection{Physical quantities}

As shown above, we have for the particle number, $\hat{n}_{r,\tau}$,

\begin{eqnarray}
\langle \hat{n}_{r,\tau} \rangle  
= \left\langle  
 N^0_{r}
+ J^0_{r,\tau}  \right\rangle.
\end{eqnarray}

\noindent Then the density-density correlation function reads

\begin{eqnarray}
\left\langle n(r_1) n(r_2) \right\rangle  = 
\frac{1}{L_{\tau}}\left\langle \sum_{\tau} \left(N^{0}_{r_1} +
J^{0}_{r_1,\tau} \right) \left(N^{0}_{r_2} +
J^{0}_{r_2,\tau} \right)\right\rangle,
\end{eqnarray}

\noindent and the structure factor, $S({\vec k})$, is its Fourier
transform.

We now focus on the canonical case. The energy is given by $ \langle E
\rangle = - \frac{\partial}{\partial\beta} \ln Z$ and therefore by

\begin{eqnarray}
\langle E \rangle  = \frac{1}{\beta} \left\langle \delta V_1 N_{V_1} + \delta V_2 N_{V_2} 
- N_H  \right\rangle  .
\end{eqnarray}

\noindent The meaning of the three terms is transparent. The first two 
are the near and next near neighbor potential energies, and the last,
which comes from the jumps, is the kinetic energy.

The superfluid density is related to the winding number, $W$, by
\cite{batrouni1,batrouni2}

\begin{equation}
\rho_s = \frac{\left\langle W^2 \right\rangle}{2t\beta} .
\end{equation}

\noindent where $W$, given by $\sum_{\tau} N^1_{\tau} $,
 changes with the global spatial currents $N^1_{\tau}$.  

Finally, we are able to calculate easily the Green function, {\it
i.e.} the correlation function of the order parameter,

\begin{eqnarray}
 G(L)= \frac{1}{2\, L_{\tau} L_x} \left\langle \sum_{r,\tau} \left(
 a_{r,\tau} a^{\dagger}_{r+L,\tau}  
 + a_{r+L,\tau} a^{\dagger}_{r,\tau}\right)\right\rangle.
\end{eqnarray}

\noindent which, in isotropic cases, equals

\begin{eqnarray}
G(L) = \frac{1}{L_{\tau} L_x}  \left\langle
 a_{r,\tau} a^{\dagger}_{r+L,\tau}  \right\rangle.
\end{eqnarray}

\noindent This can be calculated directly from the generating
functional, Eq.~31,

\begin{eqnarray}
G(L) &=&\frac{1}{2\,L_{\tau}L_x}\frac{\partial}{\partial
\left(\frac{\textstyle
\delta^L t^L }{\textstyle L\,!}\right)} \ln Z,
\end{eqnarray}

\noindent which can be easily shown to equal

\begin{eqnarray}
G(L) &=& \left\langle \frac{1}{2\,L_{\tau}L_x}
\frac{L\,!}{\left(\delta t\right)^L} N_L
\right\rangle. 
\end{eqnarray}

\noindent We now see the importance of keeping multiple jump terms:
The only configurations that contribute to $G(L)$ are those that
contain jumps of length $L$. This is physically satisfying: There are
correlations between two sites if they are directly connected by a
jump. This quantity is also simple to calculate, we just need to count
the number of multiple jumps on the lattice.

\section{Algorithmic details}

\subsection{Evolution of the system}

We begin by choosing an initial configuration of Bosons.  This is
done by introducing a current line for each particle {\it i.e.} by
putting the corresponding global time currents to 1. Then the
simulation proceeds by modifying these current lines using a
Metropolis sampling scheme.

\begin{figure}
\epsfxsize=3.0in
\epsfysize=2.5in
\epsffile{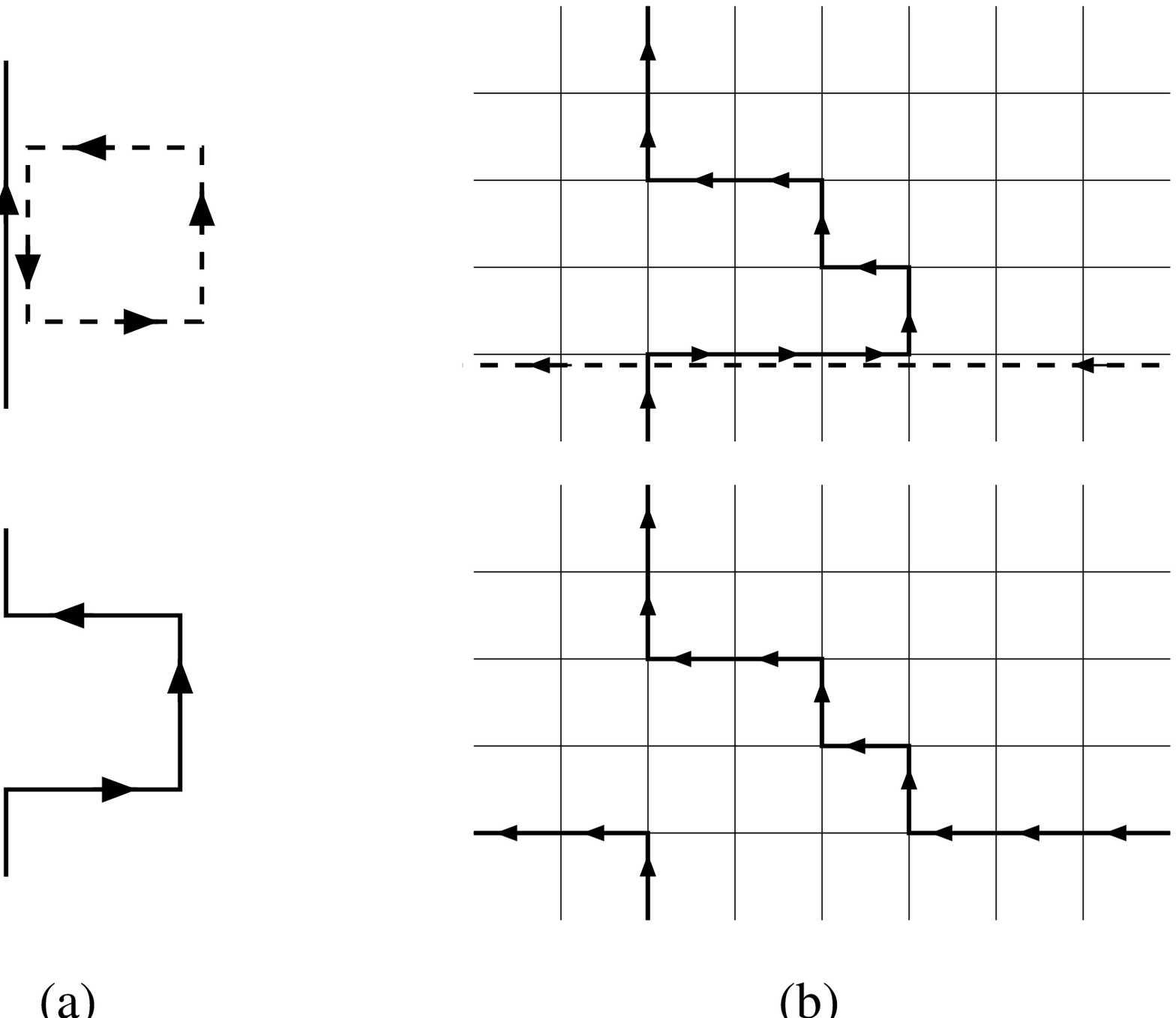}
\vskip 0.2cm
\caption{The two kinds of trial moves for the one dimensional model:
Local moves (a) and global moves (b).}
\end{figure}

In one dimension, we can explore the whole phase space by simply
applying two kinds of trial moves. First, we introduce local moves
(Fig.~2(a)): We visit sites sequentially and try to add a small
(clockwise or anticlockwise) current loop.  If this does not introduce
currents greater than 1 or negative time currents, we accept or reject
probalistically according to the Boltzmann weight, Eq.~31.  In
addition to these local moves, we have global ones (Fig.~2(b)) where
we change the value of a spatial global current traversing the
system. These moves change the winding number.  The global moves have
very low acceptance probability, which means that in practice, one
cannot efficiently sample all the winding number sectors.

In order to do simulations in the grand canonical ensemble, one can
introduce global moves along the time direction which will add or
remove particles.  One can also use the chemical potential term to
introduce a disordered external potential by imposing a site dependent
$\mu$.

\begin{figure}
\epsfxsize=3.0in
\epsfysize=2.5in
\epsffile{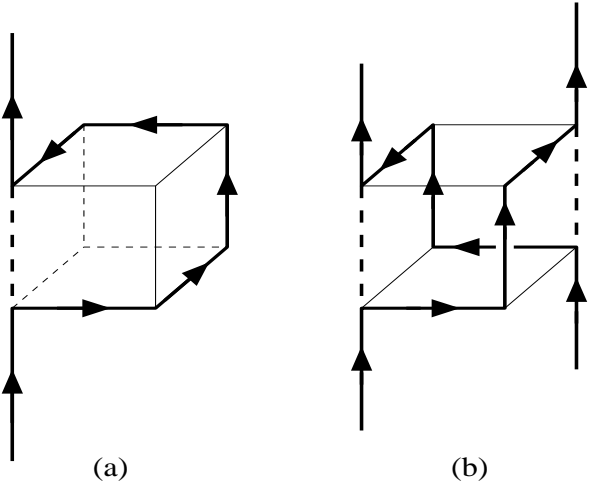}
\vskip 0.2cm
\caption{Additional trial moves for the two dimensional case: 
One-particle twist (a) and two-particle exchange (b).}
\end{figure}

In two dimensions, we need to introduce two more types of trial moves
to get topologically different configurations and ensure ergodicity.
These moves are the one particle twist and two particle exchange shown
in Fig.~3.

\begin{figure}
\epsfxsize=3.0in
\epsfysize=2.5in
\epsffile{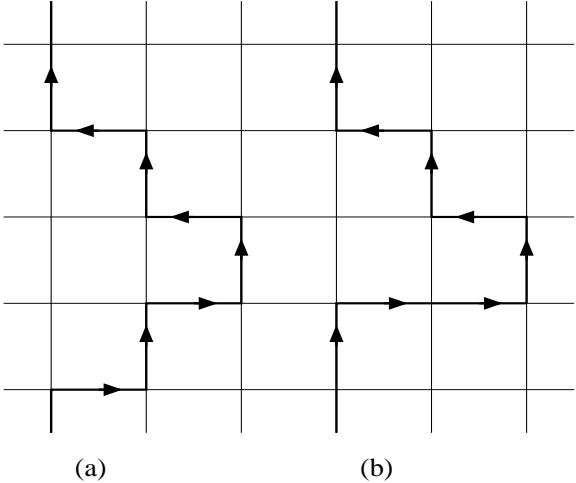}
\vskip 0.2cm
\caption{A configuration of the conserved current for one boson (a)
with four single jumps and (b) with one 
double jump and two single ones.} 
\end{figure}

\subsection{Reweighting}

As we have seen, multiple hopping terms are necessary for the
calculation of the Green function, $G$. Consider the two
configurations of Fig.~4, (a) contains only single jumps and has a
weight $(\delta t)^4$, (b) contains one double jump and two single
ones and so its weight is $(\delta t)^4/(2!)$. Both of them give a
contribution of the same order in $\delta$ to $Z$. In the first case,
since the position of each of the four jumps is arbitrary, the total
number of such configurations grows as $L^4_{\tau}$. Consequently,
they give a finite contribution of order $L^4_{\tau} (\delta t)^4=
(\beta t)^4$.  For (b), only three jumps have arbitrary positions,
which gives a final contribution of $L^3_{\tau} (\delta t)^4= (\beta
t)^3 \delta t$.  Taking the $\delta \rightarrow 0$ limit, we see that
the configurations with multiple jumps give no contribution to the
partition function.

For the same reason, all configurations with jumps of size $L+1$ and
many jumps of size $L$, do not give a finite contribution to
$G(L)$. The only important configurations for $G(L)$ are those with
one jump of size $L$ and $L$ single jumps.

Sampling the configuration with the Boltzmann weight (Eq.~31), the
probability of having multiple jumps falls to zero as $\delta$ gets
smaller.  This causes a problem: We want small $\delta$ in order to
reduce the Trotter discretization error which is proportional to
$\delta$. But in that limit the number of multiple jumps falls
exponentially making it extremely difficult to calculate the Green
function $G$.

To overcome this problem and calculate Green function efficiently, we
introduce a reweighting scheme. We replace the weight $\mathcal{P}$ by
a weight $\mathcal{P}_R$ designed to enhance configurations with
multiple jumps.

Equations 31 and 42 show that a natural choice is to keep ${\mathcal
P}_{R} = {\mathcal P}$ if there is no jump.  When there are jumps we
take

\begin{eqnarray}
{\mathcal P}_{R} = \frac{{\mathcal P}} {\left[(\delta t)^{{\mathcal
M}-1}/{\mathcal M}\, !\right]}
\end{eqnarray}

\noindent where ${\mathcal M}$ is the length of the biggest jump on 
the lattice.  Since this changes the Boltzmann weight, we should
compensate when measuring physical quantities.  For the energy, we get
(the sums run over all allowed configurations of the current)

\begin{eqnarray}
\langle E \rangle  &=& \frac{1}{\beta} \frac{\sum\left(\delta V_1 N_{V_1} 
-  N_H \right){\mathcal P}}{\sum {\mathcal P}  }  \\
&=& \frac{1}{\beta} \frac{\sum\left(\delta V_1 N_{V_1} - N_H
\right)\frac{{\mathcal P}}{{\mathcal P}_R} {\mathcal P}_R} {\sum
\frac{{\mathcal P}}{{\mathcal P}_R} {\mathcal P}_R } \nonumber\\
&=&  \frac{1}{\beta} \frac{\sum\left(\delta V_1 N_{V_1}
 -  N_H \right)\left[(\delta t)^{{\mathcal M}-1}/{\mathcal M}\,!
\right] {\mathcal P}_R}
  {\sum \left[ (\delta t)^{{\mathcal M}-1}/{\mathcal M}\,!\right]
  {\mathcal P}_R } \nonumber \\
&=& \frac{1}{\beta}\frac{\left\langle \left(\delta V_1 N_{V_1} - N_H
\right)\left[ (\delta t)^{{\mathcal M}-1}/{\mathcal M}\,!\right]
\right\rangle_{{\mathcal P}_R } } {\left\langle \left[(\delta
t)^{{\mathcal M}-1}/{\mathcal M}\,!\right] \right\rangle_{{\mathcal
P}_R } },
\nonumber
\end{eqnarray}

and for the correlation function

\begin{eqnarray}
G(L) = \frac{1}{2\ L_x \beta t} \frac{L\,!}{(\delta t)^{L-1}} 
\frac{\left\langle \left[(\delta t)^{{\mathcal M}-1}/{\mathcal M}\,! \right] 
 N_L \right\rangle _{{\mathcal P}_R }}
{\left\langle   \left[(\delta t)^{{\mathcal M}-1}/{\mathcal M}\,!\right]     
\right\rangle_{{\mathcal P}_R }. }
\end{eqnarray}

This new reweighted algorithm increases dramatically the efficiency of
the calculation of $G$. Figure 5 shows the number of jumps as a
function of the length of the jump using the original and reweighted
algorithms.  As one can see, it falls exponentially with length in both
cases.  But, for the reweighted algorithm the slope is much weaker
allowing frequent long jumps.

\begin{figure}
\epsfxsize=3.0in
\epsfysize=2.25in
\epsffile{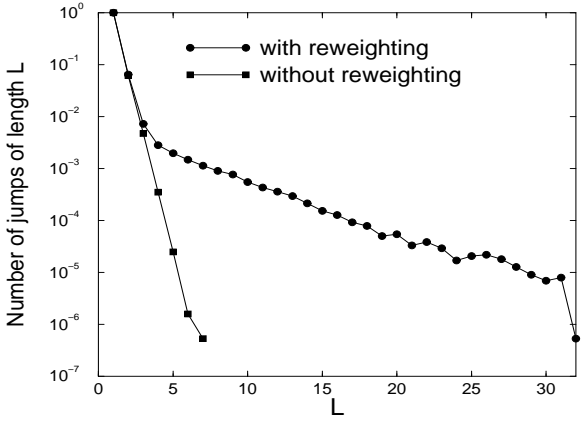}
\vskip 0.2cm
\caption{Number of jumps of different lengths with and without
reweighting for a system of 16 bosons on 32 sites at $\beta = 4$.}
\end{figure}

Reweighting favors long jumps which we can exploit to change the
global spatial currents and thus the winding number.  This algorithm
then gives us exact results (after extrapolation to $\delta = 0$) even
for small size systems where the effect of non-zero $W$ cannot be
neglected.

That does not mean that using the Boltzmann weight without reweighting
is of no interest. For big systems, non-zero winding numbers will have
very small effect on local quantities and if one only wants these,
there is no need for reweighting.

\subsection{Tests of the algorithm}

In order to test the algorithm, we compare its results with those of
exact diagonalization on a small one-dimensional system (4 bosons on
an 8 sites lattice at $\beta=4$ with only hard core interactions).  We
extrapolated our results to the $\delta = 0$ limit before comparing
(Fig.~6).  The systematic errors are, as expected, clearly linear in
$\delta$. The extrapolated results are in excellent agreement (less
than one percent error) with exact ones. Fig.~7 shows comparison for
density-density correlation function and $G$.

\begin{figure}
\epsfxsize=3.0in
\epsfysize=2.25in
\epsffile{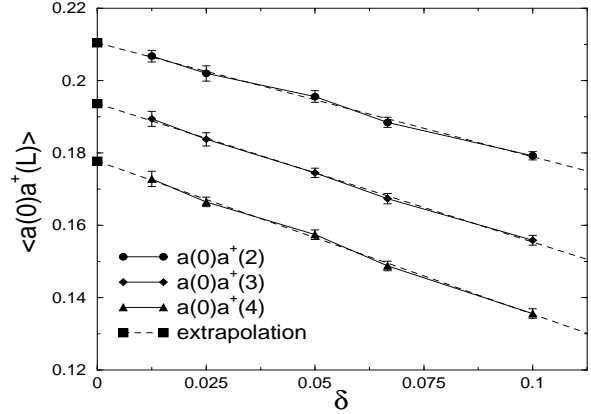}
\vskip 0.2cm
\caption{Extrapolation to $\delta=0$ of the Green function at
different distances for a system of 4 bosons on 8 sites at $\beta=4$
with only hardcore interactions.}
\end{figure}

\begin{figure}
\epsfxsize=3.0in
\epsfysize=2.25in
\epsffile{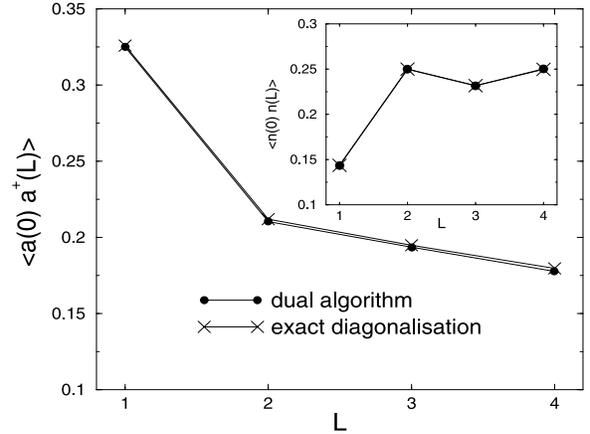}
\vskip 0.2cm
\caption{Comparison of Green function and density-density 
correlation functions obtained with the dual algorithm and exact
diagonalisation techniques. The system contains 4 bosons on 8 sites
with $\beta=4$ and only hardcore interactions.}
\end{figure}

\section{APPLICATIONS}

As discussed above, with this algorithm (as with others before it) we
can measure various physical quantities, such as the energy, the
superfluid density, $\rho_s$, the density-density correlation function
and its Fourier transform, the structure factor. What is new here is
the ease with which we can measure the Green function, $\langle
G(L)=a^{\dagger}(L) a(0)\rangle$ which, among other things, gives the
number of condensed particles (occupation of the zero momentum
mode). If one defines

\begin{eqnarray}
G(0) = \frac{1}{L_\tau L_x} \sum_{r,\tau}\langle \, a^\dagger_r a_r \rangle,
\end{eqnarray}  

\noindent then the Fourier transform of $G(L)$ equals the density
of particles in the $k$th mode. In one dimension

\begin{eqnarray}
 \frac{1}{L_x}\sum_L e^{ikL} G(L) = N(k) = \frac{\langle a^\dagger_k
 a_k \rangle}{L_x}
\end{eqnarray}

\noindent where

\begin{eqnarray} 
a^\dagger_k = \frac{1}{\sqrt{L_x}} \sum_r
e^{-ikr} a^\dagger_r.
\end{eqnarray}

\noindent Using $G(L)$ and $N(0)$, one can easily obtain valuable
information about the presence (or absence) of off-diagonal long range
order (ODLRO), \textit{i.e.}, phase coherence at long distances.

We studied with this algorithm a two-dimensional 12x12 system of
hardcore bosons at various fillings and interactions. We took
$\beta=6$, which is a low enough temperature to access ground state
properties. However, although the information is present in the
configurations generated by this algorithm, we did not measure the
Green function off-axis. We only measured it along the two lattice
axes to reduce measurement time.

\begin{figure}
\epsfxsize=3.0in
\epsfysize=2.25in
\epsffile{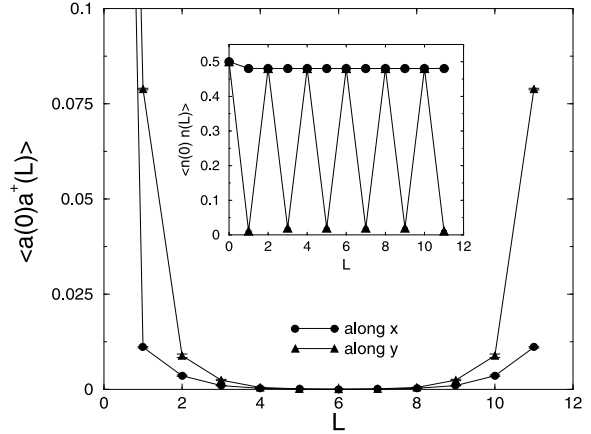}
\vskip 0.2cm
\caption{Green function and density-density correlation function 
(inset) versus distance. 72 bosons on a $12 \times 12$ lattice with
$V_1=0, V_2=3$ and $\beta=6$. The bosons form a striped incompressible
solid.}
\end{figure} 

In Fig.~8 we show results for a $12X12$ system at half filling and
strong next near neighbor repulsion, $V_1=0, V_2=3$. The nnn repulsion
makes the presence of next near neighbours too costly. Consequently,
the Bosons organize themselves into stripes with a $(\pi,0)$ (or
$(0,\pi)$) ordering vector for the structure factor. In other words,
the Bosons form an incompressible gapped solid with stripes along the
$x$ or $y$ axis. This is shown very clearly in the density-density
correlation function shown in the inset of Fig.~8. Also shown in
Fig.~8 is the Green function along the $x$ and $y$ axes. Notice that
$G(L)= \langle a(0) a^{\dagger}(L) \rangle$ falls faster along the
stripes ($x$-axis) than in the transverse direction ($y$-axis). This
is because the energy cost for transverse hops, while high, is still
finite. On the other hand, the cost for hops along the stripes is
infinite because such hops would always lead to multiple occupancy of
sites. In both cases, $G(L)$ goes to zero at long distance indicating
the absence of superfluidity. This is confirmed with direct
measurement of $\rho_s=\langle W^2\rangle/2t\beta$.

\begin{figure}
\epsfxsize=3.0in
\epsfysize=2.25in
\epsffile{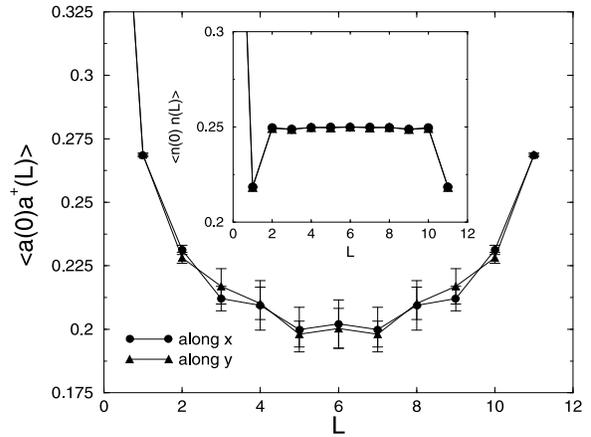}
\vskip 0.2cm
\caption{Green function and density-density correlation function 
(inset) in the superfluid phase (72 bosons on a $12 \times 12$ lattice
with $V_1 = 0, V_2=1$ and $\beta = 6$). The system is isotropic and
superfluid. }
\end{figure}

Reducing the nnn repulsion allows the long range order to melt and be
replaced by ODLRO. This is shown in Fig.~9. The inset shows the
density-density correlation functions along the $x$ and $y$ directions
to be equal. The system is therefore spatially isotropic. This is
confirmed by the isotropy (withing error bars) of $G(L)$ along $x$ and
$y$. Also, notice that $G(L)$ is finite at long distance indicating
symmetry breaking\cite{symbreak} (nonzero $\langle a\rangle $) and
thus nonzero superfluid density which we also confirmed by direct
measurement of $\rho_s = 0.304 \pm 0.017$ from the winding number
fluctuations.

\begin{figure}
\epsfxsize=3.0in
\epsfysize=2.25in
\epsffile{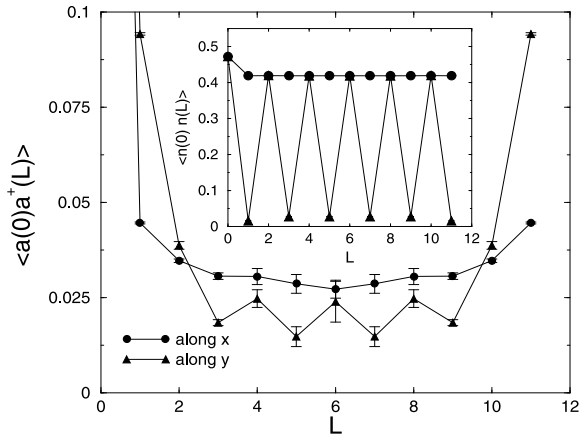}
\vskip 0.2cm
\caption{Green function and density-density correlation function 
(inset) in the supersolid phase (68 bosons on a $12 \times 12$ lattice
with $V_1=0,V_2=3$ and $\beta = 6$). The system shows at the same time
superfluidity and long range density modulations. }
\end{figure}

If, instead of reducing $V_2$ at half filling, we now repeat the
simulations that gave Fig.~8 but doped away from half filling,
$\rho=0.472$, we observe more structure developing in the Green
function. This is shown in Fig.~10. As before, the inset shows the
density-density correlation function along $x$ and $y$. As in Fig.~8,
this shows clearly the presence of stripe order along the
$x$-axis. However we also see that $G(L)$ does not decay to zero at
long distance. Instead, along the $x$-axis, $G(L)$ resembles that in
Fig.~9, {\it i.e.} tending towards a nonzero value at long distance
indicating ODLRO and superfluidity. This is again confirmed by
measuring directly the fluctuations of the winding number giving
$\rho^x_s=0.027 \pm 0.003$. Along the $y$-axis, $G(L)$ shows an
oscillatory structure: At long distances, $G(L={\rm even})$ is greater
than $G(L={\rm odd})$ and is almost equal to $G(L)$ along the
$x$-axis. This shows clearly that there is superfluidity along the
$y$-axis, transverse to the direction of the stripes. Direct
measurement gives $\rho_s^y=0.015 \pm 0.005$ which is less than
$\rho^x_s$ as may have been expected. The fact that both, long range
density wave structure (diagonal order) and superfluidity (ODLRO)
exist at the same time shows this phase to be a striped
supersolid. This phase has been shown to have a finite critical
velocity\cite{batrouni2} and not to phase separate\cite{batrouni3}. It
is therefore stable. Doping above half filling will yield similar
results: This model has particle-hole symmetry around $\rho=1/2$.

The wave structure in $G(L)$ along the $y$-axis is easy to interpret
physically. When the system was at half filling, the Bosons could not
hop along the stripes (since no double occupancy is allowed) and could
only hop with difficulty transverse to the stripes into the empty
sites due to the high energy cost. When the system is doped below half
filling, the Bosons can easily hop {\it along} the stripes into
neighboring holes. They still cannot easily hop an odd number of
lattice spacings into the mostly empty stripes due to the high cost of
nnn interactions. However, by hopping an even number of lattice
spacings, they can fill in holes in the Boson stripes, thus aquiring
near neighbors at zero potential energy cost and avoiding next near
neighbors which cost a lot of potential energy. This explains the
smaller values of $G(L)$ for odd $L$ along the $y$-axis.

This also brings out an important point. The interpretation of a
striped supersolid is {\it not} simply that the stripes form channels
along which the Bosons delocalize and the superfluid flows. While this
does indeed happen, Fig.~10 shows clearly that there is also
delocalization and superfluidity in the direction {\it transverse} to
the stripes.

\renewcommand{\thefigure}{11a}

\begin{figure}[h*]
\epsfxsize=3.0in
\epsfysize=2.25in
\epsffile{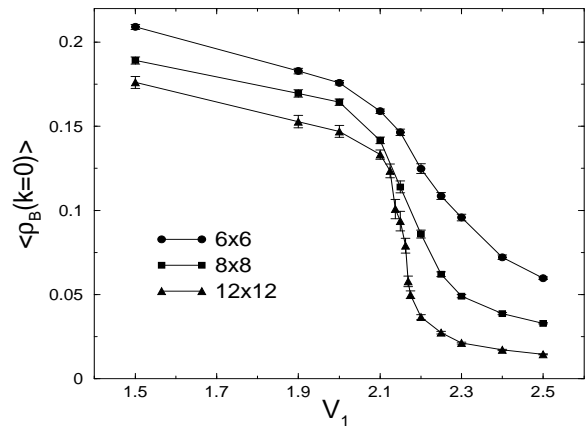}
\vskip 0.2cm
\caption{The filling fraction of the $k_x=0$ mode versus the 
interaction $V_1$ ($V_2=0$) for different sizes. The systems are half
filled and $\beta=6$. }
\end{figure}  

\renewcommand{\thefigure}{11b}

\begin{figure}[h*]
\epsfxsize=3.0in
\epsfysize=2.25in
\epsffile{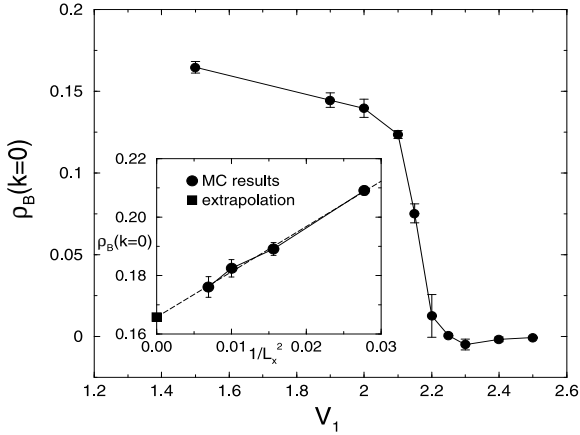}
\vskip 0.2cm
\caption{The filling fractions of Fig.~11a extrapolated to infinite
systems.  Inset: Filling fraction versus inverse system size for
$V_1=1.5$ }
\end{figure}  

\setcounter{figure}{11}
\renewcommand{\thefigure}{\arabic{figure}}

We now consider the case $V_1\ne 0, V_2=0$. As is well
established\cite{batrouni1,batrouni2,batrouni3}, for $V_1>2, \rho=0.5$
the Bosons are organized into an incompressible checkerboard solid,
while for $V_1<2$ they delocalize into a superfluid. In other words,
for large $V_1$, the Green function decays rapidly to zero as in
Fig.~8 (but is isotropic in $x$ and $y$), while for small $V_1$ it is
nonzero for large separations signalling spontaneous symmetry
breaking.

Measuring $G(L)$ as a function of $V_1$ enables us to calculate the
filling fraction of the modes, $\langle
a^{\dagger}(k_x,k_y)a(k_x,k_y)\rangle$ (Eq.~47), as a function of
interaction. Strictly speaking, this requires
$G(|x-x^{\prime}|,|y-y^{\prime}|)$, {\it i.e.} off-axis as well as
on-axis contributions to $G$. But, as explained above, we measured
only on-axis contributions to $G$. We use this to obtain an estimate
of the filling fraction as follows: Since we know that $G$ is
isotropic, and since we measured it along the lattice axis, we use
these values to interpolate to lattice separations which we have not
measured. For example, obtain $G(\sqrt{2})$, which corresponds to nnn,
we interpolate the values of $G(1)$ and $G(2)$. The resulting Green
function is Fourier transformed (two dimensional) to obtain the
filling fraction. Figure 11a shows this filling fraction for the
$k_x=0$ mode, $\rho_B(k_x=0)$, for different lattice sizes. We see a
rapid transition around $V_1=2.1$ from a phase where the condensate
fraction is large ($V_1<2$) to a phase where it is very small
($V_1>2$).  The inset of Fig.~11b shows that the extrapolation of
$\rho_B(k_x=0)$ to the infinite system is linear in the inverse system
size. The extrapolated values are shown in Fig.~11b exhibiting clearly
the phase transition from non-zero to zero condensate
density.\cite{extrap} Figure 12 shows $\rho_s$ versus $V_1$ as
directly measured from the winding number. We see that the behaviour
is exactly the same as for the condensate fraction, $N(k_x=0)$, in
Fig.~11. The transition of $\rho_s$ to zero goes hand in hand with
that of the condensate fraction.  These results for $\rho_s$ and the
zero mode filling fraction show that the transition from superfluid to
$(\pi,\pi)$-solid is continuous when $V_2=0$.

The question of what happens when this system is doped away from
$\rho=0.5$ is not simple for large $V_1$ and has been addressed
elsewhere.\cite{batrouni3}

So far we have discussed the two limits $(V_1=0, V_2\ne 0)$ and
$(V_1\ne 0, V_2=0)$. To construct the phase diagram in the $(V_1,V_2)$
plane we need to do simulations with both interactions non-zero. This
was done in Ref.~12 for $V_1 \le 4.6, V_2\le 3$ where it was found
that within these values of the interactions there was no direct
$(\pi,\pi)$-solid to $(\pi,0)$-solid transition although mean field
calculations predict this to happen starting at $(V_1=4,
V_2=2)$.\cite{batrouni2} Here we extend the simulations
farther. Figure 13 shows $\rho_s, S(\pi,\pi), {\rm and } S(\pi,0)$ as
a function of $V_2$ for $V_1=6$. We see that even at this large value
of $V_1$ there is no direct solid-solid transition. In addition, the
sharpness of the superfluid to $(\pi,0)$-solid transition suggests
that it is first order while the relative smoothness of the superfluid
to $(\pi,\pi)$-solid suggests that it is continuous. This is in
agreement with Refs.~12 and 19. Preliminary results indicate that even
at $V_1=8$ there still is no solid to solid transition. Apparently,
the frustration caused by the competition between $V_1$ and $V_2$
prevents this and always results in a superfluid phase separating the
two solid phases for any large but finite $V_{1,2}$.

One of our goals in searching the phase diagram is to explore the
possibility of a normal state at zero temperature. This is a
particularly interesting question since such states have been found
experimentally in high $T_c$ superconducotrs. Explanations offered so
far rely on the idea that the charge carriers in this normal state are
the Cooper pairs, and thus Bosonic. However, the models which have
been examined are for soft core Bosons at high densities. Our
simulations for hard core Bosons have not shown any candidates for
such a normal state. All phases we found so far are clearly solid,
superfluid or solid as shown, for example, in Fig.~13.

\begin{figure}
\epsfxsize=3.0in
\epsfysize=2.25in
\epsffile{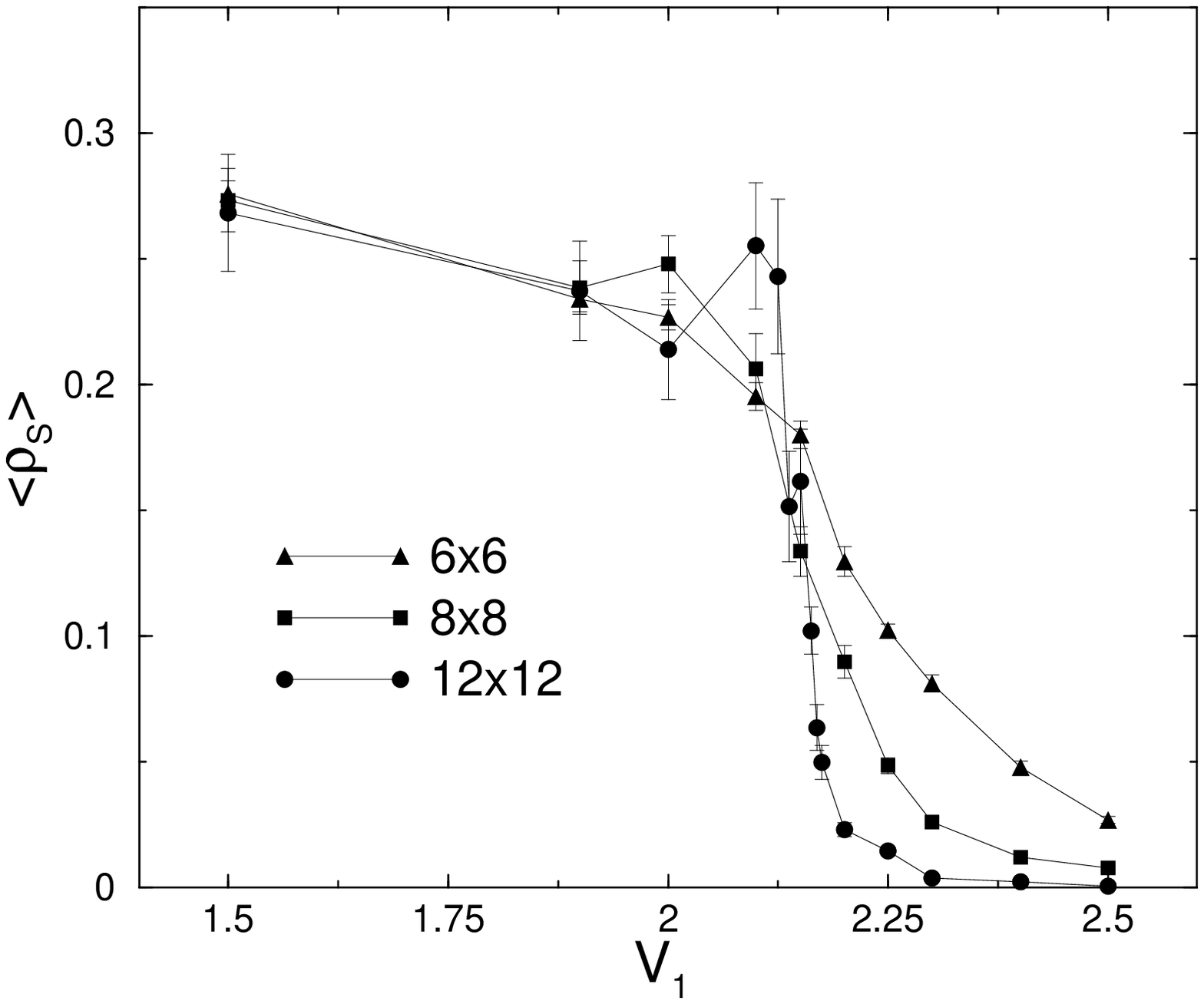}
\vskip 0.2cm
\caption{The superfluid density versus the interaction $V_1$ ($V_2=0$)
and for different sizes. The systems are half filled and $\beta = 6$.}
\end{figure}

\begin{figure}
\epsfxsize=3.0in
\epsfysize=2.25in
\epsffile{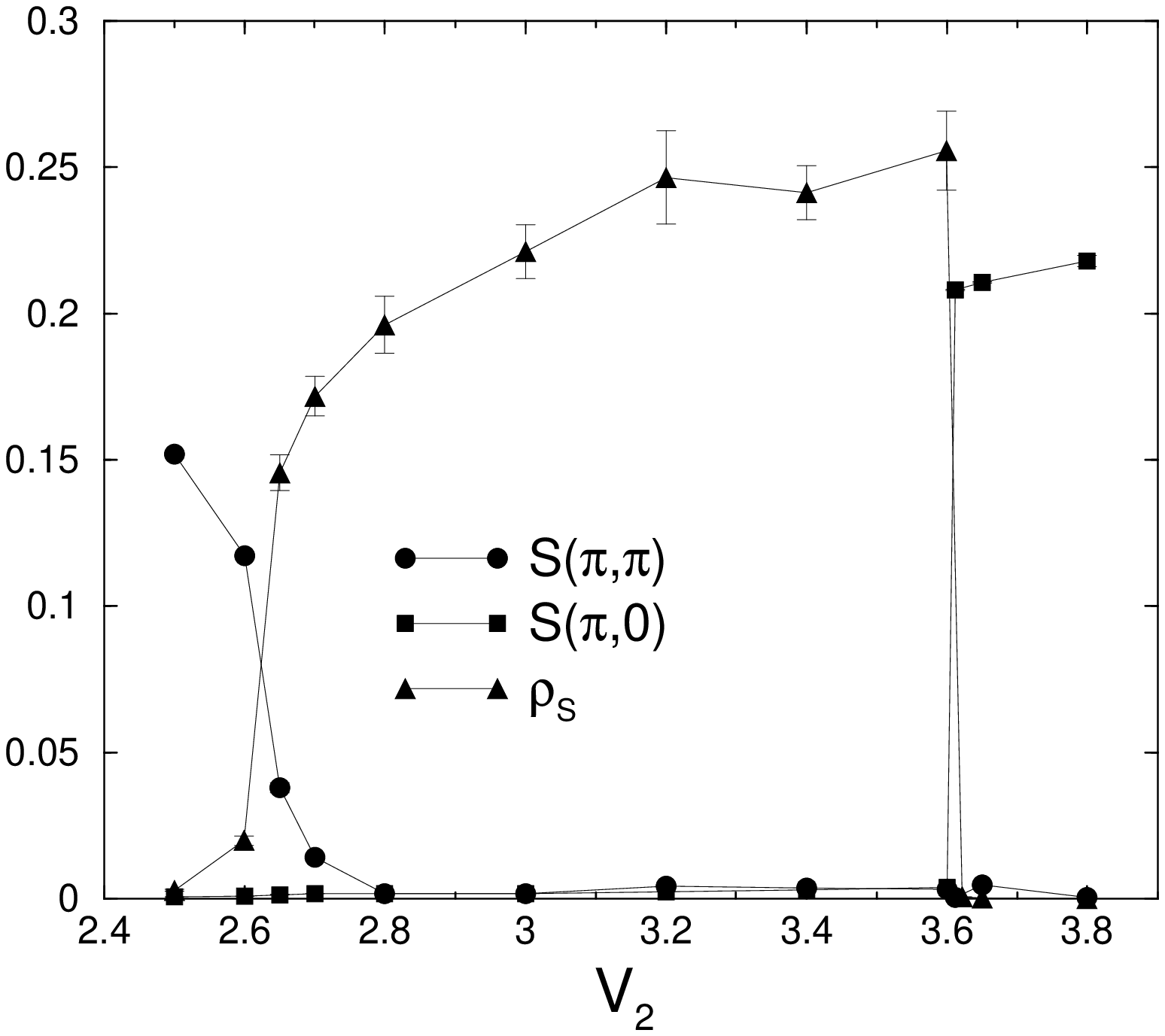}
\vskip 0.2cm
\caption{$S(\pi,\pi), S(\pi,0) {\rm and} \rho_s$ versus $V_2$ for
$V_1=6$. There is no solid-solid transition even at this large value
of $V_1$.}
\end{figure}

\section{conclusion}

We have detailed the construction and testing of a Quantum Monte Carlo
algorithm based on the exact duality transformation of the Bosonic
Hubbard model. In particular we showed that this formulation yields a
particularly simple dual partition function in the very important case
of hardcore Bosons. We also showed how to measure various physical
quantities. In particular, we showed how to measure the Green
function, an important quantity which is particularly difficult to
obtain with other algorithms. To improve the statistics of this
quantity we enhanced the efficiency of the algorithm by introducing a
new reweighting procedure.

We then used this algorithm to elaborate the structure of the various
ground state phases of the Bosonic Hubbard model. In particular, we
showed and interpreted the behaviour of the Green function in the
various superfluid, superolid and solid phases and calculated the
filling fraction in the zero momentum mode. We found a filling
fraction of about $33\%$ at weak nn coupling, $V_1=1.5$. This value is
consistent with the pure hardcore value of Ref.~21 (between $36\%$ and
$22\%$). However, the two results were obtained at different Boson
densities and at the moment it is not clear how our value depends on
the density. This is in progress.

We confirmed that the order of the phase transition from the
superfluid to the solid phases depends on the long range order.  At
half filling, the SF-$(\pi,\pi)$-solid transition is second order
while the SF-$(\pi,0)$-solid transition is first order.

The presence of a direct solid-solid transition from $(\pi,\pi)$ to
$(\pi,0)$ is still an open question although the indications are that
it is not present at any finite value of $V_{1,2}$. Another important
issue is the possibility of a normal conducting phase. Several
different scenarios for the existence of normal conducting states in
two dimensions have been offered but for softcore Bosons. So far, our
exploration of the phase diagram of this hardcore Bosonic model have
given no hints of the presence of such a state.

\medskip

\centerline{\textbf{Acknowledgments}}

We acknowledge very fruitful discussions with H.~Rieger,
R. ~T. ~Scalettar.  G.G.B.~thanks the Physics Department at U.C.~Davis
for its hospitality.

\appendix

\section{derivation of the dual partition function}
 
We present in some detail the integration of the original fields to
obtain the dual expression for Z. Then we show, using hardcore bosons
as an example, how to simplify this expression by explicitly solving
constraints appearing in it.

\subsection{General partition function}

To integrate out the bond variable $\Theta$ and the amplitude $\rho$
in Eq. 29, we expand the exponentials of complex exponentials in
Taylor series. This results in new integer variables $l_{r,\tau}$,
$p^{(i)}_{r,\tau}$ and $q^{(i)}_{r,\tau}$. $l$ arises from the
expansion of the second exponential in Eq. 29, $p^{(i)}$ and $q^{(i)}$
  from the $i^{th}$ hopping term.

\begin{eqnarray}
Z &=& \sum \int \prod_{r,\tau} \Biggl[ \rho_{r,\tau} 
\frac{d\rho_{r,\tau}}{\pi}\,
\tilde{\rho}_{r,\tau} \frac{d\tilde{\rho}_{r,\tau}}{\pi}
d\Theta^1_{r,\tau} d\Theta^0_{r,\tau} d\tilde{\Theta}^0_{r,\tau}
\nonumber \\ &\times& e^{ \left(- \rho_{\scriptstyle
r,\tau}^{\scriptstyle 2} -
\tilde{\rho}_{\scriptstyle r,\tau}^{\scriptstyle 2}     \right)} 
e^{-\delta V(n_{\scriptstyle r,\tau})} \\ &\times& \frac{ (\delta
t)^{p_{\scriptstyle r,\tau}^{\scriptstyle (1)} + q_{\scriptstyle
r,\tau}^{\scriptstyle (1)}} (\delta^2 t^2/2!)^{ p_{\scriptstyle
r,\tau}^{\scriptstyle (2)} + q_{\scriptstyle r,\tau}^{\scriptstyle
(2)}}
\cdots }
{n_{r,\tau}! \,
l_{r,\tau}!\,p^{(1)}_{r,\tau}!\,q^{(1)}_{r,\tau}!\,p^{(2)}_{r,\tau}!\,
q^{(2)}_{r,\tau}!\cdots} \nonumber \\ &\times&
\tilde{\rho}_{r,\tau}^{\textstyle{ \bigl[\, n_{r,\tau} + l_{r,\tau-1}
+ p_{r-1,\tau-1}^{(1)} + q_{r,\tau-1}^{(1)} }} \nonumber\\ && \qquad
\qquad \qquad^{\textstyle{+p_{r-2,\tau-1}^{(2)} + q_{r,\tau-1}^{(2)} +
\cdots \ \bigr]}} \nonumber \\ &\times& \rho_{r,\tau}^{\textstyle{
\bigl[\, n_{r,\tau} + l_{r,\tau} + p_{r,\tau}^{(1)} +
q_{r-1,\tau}^{(1)} }} \nonumber\\ && \qquad \qquad
\qquad^{\textstyle{+p_{r,\tau}^{(2)} + q_{r-2,\tau}^{(2)} + \cdots \
\bigr]}} \nonumber \\ &\times & e^{\textstyle -i\,
\tilde{\Theta}^{0}_{r,\tau} \bigl[ \, l_{r,\tau} + p_{r-1,\tau}^{(1)}
+ q_{r,\tau}^{(1)} + p_{r-2,\tau}^{(2)} +}\nonumber \\ && \hspace{3cm}
^{\textstyle + q_{r,\tau}^{(2)} + \cdots -J^{0}_{r,\tau} - N^{0}_{r}
\bigr]}\nonumber \\ &\times & e^{\textstyle -i\, \Theta^{1}_{r,\tau}
\bigl[ \, p_{r,\tau}^{(1)} - q_{r,\tau}^{(1)} + p_{r,\tau}^{(2)} +
p_{r-1,\tau}^{(2)} - q_{r,\tau}^{(2)} - }\nonumber \\ && \hspace{3cm}
^{\textstyle- q_{r-1,\tau}^{(2)} + \,\cdots\, -J^{1}_{r,\tau} -
N^{1}_{r,\tau} \bigr]}\nonumber \\ &\times& e^{\textstyle -i\,
\Theta^{0}_{r,\tau} \bigl[ n_{r,\tau} - J^0_{r,\tau-1} -N^0_r\bigr]}
\Biggr] \nonumber 
\end{eqnarray}

\noindent where the sum runs over all the discrete indices we have 
introduced: $l$, $p$, $q$, $J$, $N$, and $n$. Integrating over $\Theta^1,
\Theta^0 {\rm and} \tilde{\Theta}^0$ will give three constraints on the 
dual variables enforced by Kronecker deltas. The constraint coming
 from the integration over $\Theta^{0}_{r,\tau}$, enforces the
condition

\begin{eqnarray}
J^0_{r,\tau-1} + N^0_r= n_{r,\tau}.
\end{eqnarray}

\noindent The total conserved current $J^\mu_{r,\tau} + 
N^\mu_{r,\tau}$ is then just the current of bosons crossing the
lattice. Negative time currents are not allowed: Bosons cannot go
backward in imaginary time.

Using the two remaining constraints, one can see that the power of
$\tilde{\rho}_{r,\tau}$ is $2 (J^0_{r,\tau-1}+N^0_r)$. The integration
over $ \tilde{\rho}$ then yields a $(J^0_{r,\tau} + N^0_r)!$ which
compensates the $1/n_{r,\tau}!= 1/(J^0_{r,\tau} + N^0_r)!$ already
present in the expression.

The remaining $\rho$ integrals are gaussian and can be performed
easily. This leads to the dual formulation of $Z$:

\begin{eqnarray}
Z &=& \sum \prod_{r,\tau} \Biggl[ \, e^{-\delta V(J^{\scriptstyle
0}_{\scriptstyle r,\tau}+ N^{\scriptstyle 0}_{\scriptstyle r,\tau})}\\
& \times & \frac{ (\delta t)^{\bigl( p_{\scriptstyle
r,\tau}^{\scriptstyle (1)} + q_{\scriptstyle r,\tau}^{\scriptstyle
(1)}\bigr)} (\delta^2 t^2/2!)^{ \bigl( p_{\scriptstyle
r,\tau}^{\scriptstyle (2)} + q_{\scriptstyle r,\tau}^{\scriptstyle
(2)}\bigr)}
\cdots }
{l_{r,\tau}!\,p^{(1)}_{r,\tau}!\,q^{(1)}_{r,\tau}!\,p^{(2)}_{r,\tau}!\,
q^{(2)}_{r,\tau}!\cdots} \nonumber \\ &\times& \Gamma \biggl( 1
+\frac{1}{2}\Bigl[ l_{r,\tau} + l_{r,\tau+1} +p_{r-1,\tau}^{(1)} +
p_{r,\tau+1}^{(1)} + \nonumber \\ & &\hspace{3cm} + q_{r,\tau}^{(1)} +
q_{r-1,\tau+1}^{(1)} + \cdots\Bigr] \biggr) \nonumber \\ &\times &
\delta \bigl( \, l_{r,\tau} + p_{r-1,\tau}^{(1)} + q_{r,\tau}^{(1)} +
p_{r-2,\tau}^{(2)} +\nonumber \\ && \hspace{3cm} + q_{r,\tau}^{(2)} +
\cdots -J^{0}_{r,\tau} - N^{0}_{r} \bigr)\nonumber \\ &\times &
\delta\bigl( \, p_{r,\tau}^{(1)} - q_{r,\tau}^{(1)} + p_{r,\tau}^{(2)}
+ p_{r-1,\tau}^{(2)} - q_{r,\tau}^{(2)} - \nonumber \\ && \hspace{3cm}
- q_{r-1,\tau}^{(2)} + \,\cdots\, -J^{1}_{r,\tau} - N^{1}_{\tau}
\bigr)\nonumber
\Biggr]
\end{eqnarray}
 
where the sum runs over all the remaining integer indeces $l$, $p$,
$q$, and the conserved current $J+N$.

It is impossible to construct an efficient Monte Carlo algorithm using
this equation: There are many different fields, and many constraints.
If one proposes a trial move, it will be rejected most of the time
because it does not respect the conditions imposed by the $\delta$'s
or the factorials.

Of course the role of the $\delta$-functions is to allow only the
acceptable configurations and give them the correct weight. It is
possible to simplify greatly this partition function by solving these
constraints, \textit{i.e.} by finding the allowed configurations of
currents and the corresponding weights. We now show how to do this for
the case of hardcore bosons.

\subsection{Hardcore bosons}

To impose the hard core limit, we require the number of particles not
be greater than 1 on any site. This is achieved simply by imposing
that $J^{0}_{r,\tau} + N^{0}_r=0,1$ and $| J^1_{r,\tau} +
N^1_{r,\tau}|=0,1$ on all bonds. It is also understood that the
currents must always be conserved.

Now we need to find the Boltzmann weights of the allowed
configurations. To do this we calculate the weights of representative
Boson ({\it i.e.} current) configurations.  The contributions of the
interaction and chemical potential are very simple to include. We have
an $\exp(\delta \mu)$ factor for each non-zero time current element,
and an $\exp(-\delta V_1)$ for each pair of near neighbour time
currents.

\begin{figure}
\epsfxsize=3.0in
\epsfysize=2.25in
\epsffile{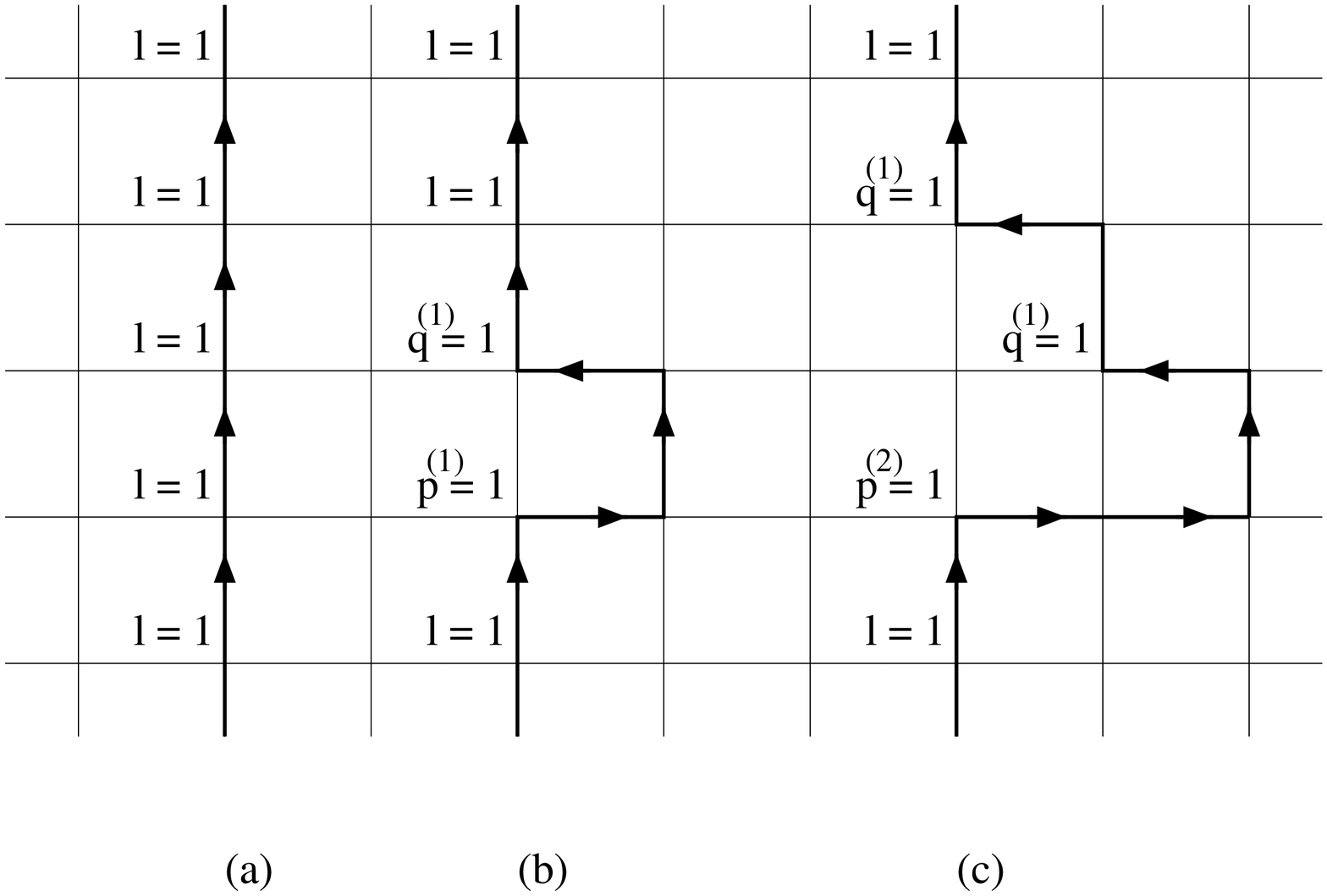}
\vskip 0.2cm
\caption{Different configurations of the conserved current. (a)
without jumps, (b) with two single jumps, and (c) with one double jump
and two single ones. The non-zero values of the integer dual variables
are indicated.}
\end{figure}

For example, consider configuration (a) in Fig.~13 where the time
current, $J^{0}_{r,\tau} + N^{0}_{r}$, is 1 along the directed path
shown and zero everywhere else. This configuration satisfies all the
constraints, and is uniquely specified by having all integer fields
equal to zero except for the $l$ field along the directed path where
it is equal to 1. This is shown in the figure. So the Boltzmann weight 
of this configuration is simply

\begin{eqnarray}
\prod_{\tau}  e^{\delta \mu} \Gamma(2) = (e^{\delta \mu})^{L_\tau} 
= e^{\beta \mu}
\end{eqnarray}

\noindent which agrees with intuitive expectations.

Now consider path (b) in Fig.~13. Clearly, this is an allowed
configuration and as before we determine, by explicit calculation, the
values of the integer fields that give it. These are shown in the
figure. The Boltzmann weight of this configuration is therefore:

\begin{eqnarray}
(e^{\delta \mu})^{L_\tau} (\delta t)^2 \left(\Gamma(2)\right)^{L_\tau}
= e^{\beta \mu} (\delta t)^2
\end{eqnarray}

The physical picture that emerges is simple: We always have the usual
chemical potential factor and when the current line jumps from one
site to another, we have a contribution of the hopping term. The $l$
are non zero where there are time currents, the $p$ where there is a
jump to the right, and the $q$ where there is a jump to the left.  As
a third example, to fix ideas, consider path (c) in Fig.~13. It has
one double jump and two single ones. Therefore, we now have one
non-zero $p^{(2)}$, corresponding to the double jump to the right, and
two $q^{(1)}$ for the two single jumps to the left. The weight of this
configuration is

\begin{eqnarray}
\frac{\delta^2 t^2}{2!} (\delta t)^2   e^{\beta \mu} 
\end{eqnarray}

Thus, in the hardcore limit, the partition function simplifies
drastically.  Only one field survives: The conserved hard core
current. The weight of a configuration can be easily calculated as
follows

\begin{eqnarray}
 Z &=& {\sum_{\{J, N\}}}' {\mathcal P}(N,J) \\ & = & {\sum_{ \{J, N \}
 }}' \prod_L \biggl[ \left(\frac{(\delta t)^L}{L!}\right)
 ^{N_{\scriptstyle l}}\biggr] e^{-\delta V_1 N_{\scriptstyle V_1}}
 e^{\beta \mu N_B} \nonumber
\end{eqnarray}

\noindent where $N_{L}$ is the number of jumps of length $L$, $N_{V_1}$ is 
the number of near-neighbor time current on the lattice, $N_b$ is
simply the number of bosons, and ${\mathcal P}$ is the Boltzmann
weight.  As mentioned in the main body of the paper, this partition
function has systematic Trotter error of order $\delta$.  If one wants
to construct an algorithm accurate to order $\delta^2$, one would have
to keep track of all the second order terms all along the development
of the kinetic term.

\begin{figure}
\epsfxsize=2.00in
\epsfysize=2.00in
\centerline{\epsffile{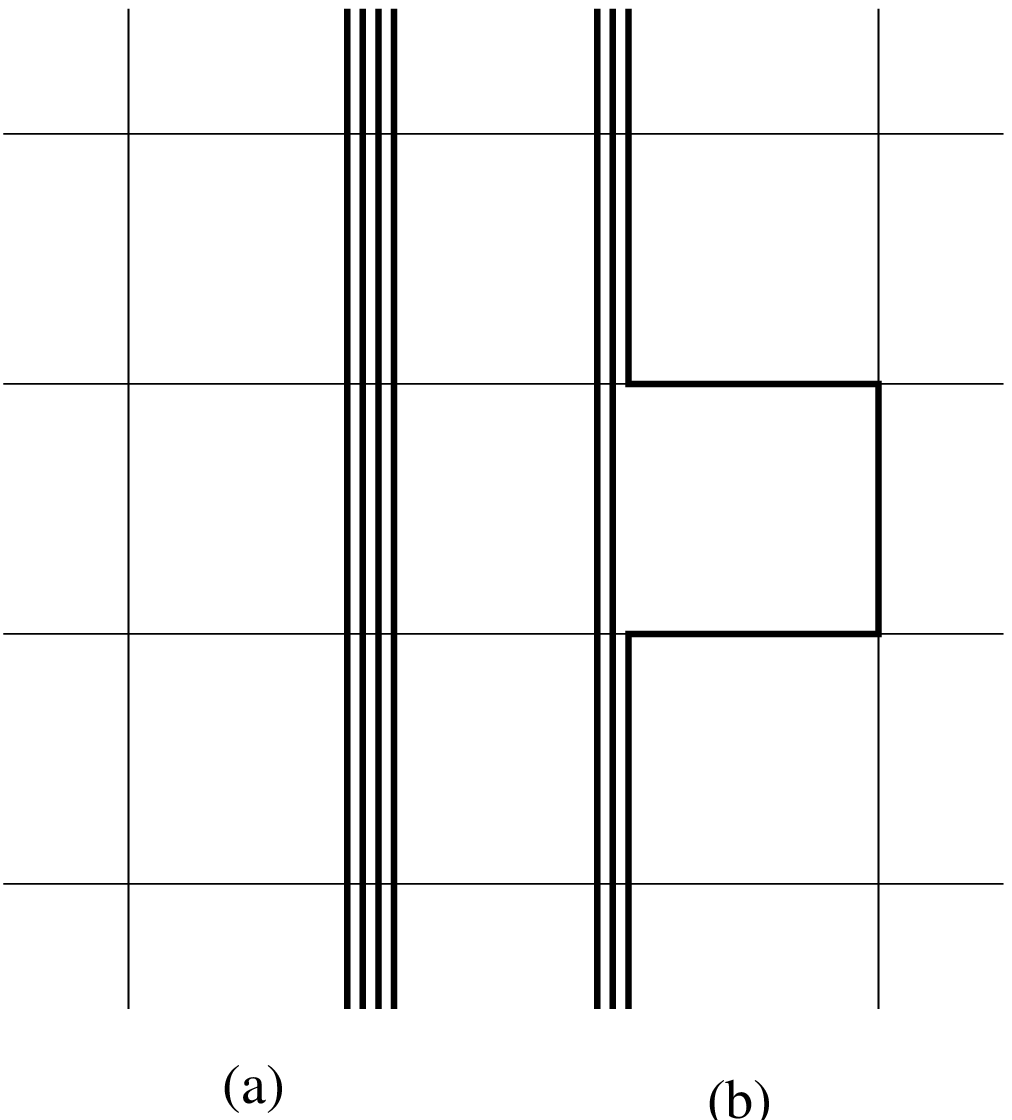}}
\vskip 0.2cm
\caption{Configuration with softcore bosons. (a) four bosons at
the same site, (b) three bosons at the same site and one of them
hopping to a neighboring site.}
\end{figure}

One can generalize this approach to other kinds of systems. For
example if we look at the soft core configurations shown in
Fig.~14(a), four bosons cross the lattice on the same site without
hopping. The only non zero field is $ l_{r',\tau}$ and we get the
weight

\begin{eqnarray} 
\left(\frac{1}{4!}\, \Gamma(5)\,e^{4\delta \mu - 6\delta V_0}
\right)^{L_{\tau}} =
 e^{\beta(4 \mu - 6 V_0)},
\end{eqnarray}

\noindent which is the expected result. Configuration (b) of Fig.~14
with three bosons on the same site and one of them hopping to a
neighboring site contributes

\begin{eqnarray}
3 \ e^{ 3\beta \mu} \ e^{ -\delta V_{\scriptstyle 0} (3L_{\scriptstyle
\tau} -2)}
\ e^{ -2 \delta V_{\scriptstyle 1}} \ (\delta t)^2.
\end{eqnarray}

\noindent The prefactor, 3, can be interpreted physically: 
Each boson can make this jump with equal probability and the sum must
be taken.


\begin{references}
\bibitem{symbreak} The symmetry breaking is, of course, not spatial.
                   It is the $U(1)$ symmetry of the Hamiltonian which
                   is broken when the system becomes superfluid.
\bibitem{extrap} It is important to keep in mind that these extrapolated 
                   values still include some Trotter error due to the
                   discretization of imaginary time. Also, although
                   the temperature is low ($\beta=6$) it is not zero,
                   this will introduce some corrections too.
\end{references}
\end{document}